\documentclass{ws-mpla}

\begin{document}
\def\gsim{\mathrel{\rlap{\lower4pt\hbox{\hskip1pt$\sim$}}
    \raise1pt\hbox{$>$}}}       
    
\markboth{Xavier Calmet}
{Xavier Calmet}

\catchline{}{}{}{}{}

\title{Planck Length and Cosmology}

\author{\footnotesize Xavier Calmet}

\address{Service de Physique Th\'eorique, CP225 \\
Boulevard du Triomphe \\
B-1050 Brussels \\
Belgium\\
xavier.calmet@ulb.ac.be}

\maketitle

\pub{Received (Day Month Year)}{Revised (Day Month Year)}

\begin{abstract}
We show that an unification of quantum mechanics and general relativity implies that there is a fundamental length in Nature in the sense that no operational procedure would be able to measure distances shorter than the Planck length. Furthermore we give an explicit realization of  an old proposal by Anderson and Finkelstein who argued that a fundamental length in nature implies unimodular gravity. Finally, using hand waving arguments we show that a minimal length might be related to the cosmological constant which, if this scenario is realized, is time dependent.

\end{abstract}

\ccode{PACS Nos.: 98.80.-k, 04.20.-q.}

\section{Introduction}	

The idea that a unification of quantum mechanics and general relativity implies the notion of a fundamental length is not new\cite{minlength1}. However, it has only recently been established that no operational procedure could exclude the discreteness of space-time on distances shorter than the Planck length\cite{Calmet:2004mp}. This makes the case for a fundamental length of the order of the Planck length much stronger. It seems reasonable to think that any quantum description of general relativity will have to include the fact that measurement of distance shorter than the Planck length are forbidden. It is notoriously difficult to build a quantum theory of gravity.  Besides technical difficulties the lack of experimental guidance, the Planck length being so miniscule $l_P\sim 10^{-33}$cm, is flagrant. In this work we shall however argue that a fundamental length in nature, even if it is as small as the Planck scale may have dramatic impacts on our universe. In particular, we will argue that it may be related to the vacuum energy, i.e. dark energy and thus account for roughly 70$\%$ of the energy of the universe.

We shall first  present our motivation for a minimal length which follows from quantum mechanics, general relativity and causality. We will then argue that a fundamental length in nature may lead to unimodular gravity. If this is the case, the cosmological constant is an integration parameter and is thus arbitrary. Finally, we shall consider argument based on spacetime quantization to argue that the cosmological constant might not be actually constant but might be time dependent. 

\section{Minimal Length from Quantum Mechanics and General Relativity}

We first review the results obtained in ref.\cite{Calmet:2004mp}. We show that quantum mechanics and classical general relativity considered simultaneously imply the existence of a minimal length, i.e. no operational procedure exists which can measure a
distance less than this fundamental length. The key ingredients used to
reach this conclusion are the uncertainty principle from quantum
mechanics, and gravitational collapse from classical
general relativity.

A dynamical condition for gravitational collapse is given by the hoop
conjecture\cite{hoop}: if an amount of energy
$E$ is confined at any instant to a ball of size $R$, where $R < E$,
then that region will eventually evolve into a black hole. We
use natural units where $\hbar, c$ and Newton's constant (or $l_P$)
are unity. We also neglect numerical factors of order one.

From the hoop conjecture and the uncertainty principle, we immediately deduce the
existence of a  minimum ball of size $l_P$. Consider a particle
of energy $E$ which is not already a black hole. Its size $r$ must
satisfy 
\begin{eqnarray}
 r \gsim {\rm \bf max} \left[\, 1/E\, ,\,E \, \right]~~,
\end{eqnarray} 
where $\lambda_C \sim 1/E$ is its Compton wavelength and $E$
arises from the hoop conjecture. Minimization with respect to $E$
results in $r$ of order unity in Planck units or $r \sim l_P$.
If the particle is a black hole, then its radius grows with mass: $r \sim E \sim 1/
\lambda_C$. This relationship suggests that an experiment designed (in
the absence of gravity) to measure a short distance $l << l_P$ will
(in the presence of gravity) only be sensitive to distances
$1/l$. 

Let us give a concrete model of minimum length. Let the position operator $\hat{x}$ have
discrete eigenvalues $\{ x_i \}$, with the separation between
eigenvalues either of order $l_P$ or
smaller. For regularly distributed eigenvalues with
a constant separation, this would be equivalent to a spatial lattice.
We do not mean to imply that nature implements minimum length in this particular
fashion - most likely, the physical mechanism is more complicated, 
and may involve,
for example, spacetime foam or strings. However, our concrete 
formulation lends itself
to detailed analysis. We show below that this formulation
cannot be excluded by any gedanken experiment, which is strong evidence for the
existence of a minimum length.

Quantization of position does not by itself imply quantization of
momentum. Conversely, a continuous spectrum of momentum does not imply
a continuous spectrum of position. In a formulation of
quantum mechanics on a regular spatial lattice, with spacing $a$
and size $L$, the momentum operator has eigenvalues which are
spaced by $1/L$. In the infinite volume limit the momentum operator can have
continuous eigenvalues even if the spatial lattice spacing is kept
fixed. This means that the displacement operator 
\begin{eqnarray} \label{disp}
\hat{x} (t) - \hat{x} (0) = \hat{p}(0) {\frac{t}{M}} 
\end{eqnarray}
does not necessarily have discrete eigenvalues (the right hand side of
(\ref{disp}) assumes free evolution; we use the Heisenberg picture
throughout). Since the time evolution operator is unitary the
eigenvalues of $\hat{x}(t)$ are the same as $\hat{x}(0)$. Importantly
though, the spectrum of $\hat{x}(0)$ (or $\hat{x}(t)$) is completely
unrelated to the spectrum of the $\hat{p}(0)$, even though they are
related by (\ref{disp}).  A measurement of arbitrarily small displacement
(\ref{disp}) does not exclude our model of minimum length. To
exclude it, one would have to measure a position eigenvalue $x$ and a nearby eigenvalue $x'$, with $|x - x'| << l_P$.

Many minimum length arguments are obviated by the simple observation of the minimum ball. However,
the existence of a minimum ball does not by itself preclude the
localization of a macroscopic object to very high precision.
Hence, one might attempt to measure the spectrum of $\hat{x}(0)$
through a time of flight experiment in which wavepackets of
primitive probes are bounced off of well-localised macroscopic
objects. Disregarding gravitational effects, the discrete spectrum
of $\hat{x}(0)$ is in principle obtainable this way. But,
detecting the discreteness of $\hat{x}(0)$ requires wavelengths
comparable to the eigenvalue spacing.  For eigenvalue spacing
comparable or smaller than $l_P$, gravitational effects cannot be
ignored, because the process produces minimal balls (black holes)
of size $l_P$ or larger. This suggests a direct measurement of the
position spectrum to accuracy better than $l_P$ is not possible.
The failure here is due to the use of probes with very short wavelength.

A different class of instrument, the interferometer,  is capable of measuring
distances much smaller than the size of any of its sub-components.  Nevertheless, the uncertainty principle and gravitational collapse prevent an arbitrarily accurate measurement of
eigenvalue spacing.  
First, the limit from quantum mechanics. Consider
the Heisenberg operators for position $\hat{x} (t)$ and momentum
$\hat{p} (t)$ and recall the standard inequality 
\begin{eqnarray} \label{UNC}
(\Delta A)^2 (\Delta B)^2 \geq  ~-{1 \over 4} ( \langle [
\hat{A}, \hat{B} ] \rangle )^2 ~~.
\end{eqnarray} 
Suppose that the position of a free mass is measured at time $t=0$
 and again at a later time.
The position operator at a later time $t$ is 
\begin{eqnarray} \label{P} \hat{x}
(t) = \hat{x} (0) ~+~ \hat{p}(0) \frac{t}{M}~~. 
\end{eqnarray} 
We assume a free particle Hamiltonian here for simplicity, but the argument can be generalized\cite{Calmet:2004mp}. The commutator between the position operators at $t=0$ and $t$
is 
\begin{eqnarray} 
 [ \hat{x} (0), \hat{x} (t)] ~=~ i {t \over M}~~,
\end{eqnarray} 
 so using (\ref{UNC}) we have 
 \begin{eqnarray}  \vert \Delta x (0) \vert
\vert \Delta x(t) \vert \geq \frac{t}{2M}~~.
\end{eqnarray} 
We see that at least one of the uncertainties $\Delta x(0)$ or $\Delta x(t)$
must be larger than of order $\sqrt{t/M}$.
As a measurement of the discreteness of $\hat{x}(0)$
requires {\em two} position measurements,
it is limited by the greater of $\Delta x(0)$ or $\Delta x(t)$,
that is, by $\sqrt{t/M}$, 
 \begin{eqnarray} 
 \label{SQL} \Delta x \equiv {\rm \bf max}\left[
 \Delta x(0), \Delta x(t) \right]
 \geq
\sqrt{ t \over 2 M }~~, 
\end{eqnarray} 
where $t$ is the time over
which the measurement occurs and $M$ the mass of the object whose
position is measured. In order to push $\Delta x$ below $l_P$, we
take $M$ to be large.  In order to avoid
gravitational collapse, the size $R$ of our measuring device must
also grow such that $R > M$. However, by causality $R$ cannot
exceed $t$. Any component of the device a distance greater than
$t$ away cannot affect the measurement, hence we should not
consider it part of the device. These considerations can be
summarized in the inequalities 
\begin{eqnarray} 
 \label{CGR} t > R > M
~~.\end{eqnarray} 
Combined with (\ref{SQL}), they require $\Delta x
> 1$ in Planck units, or 
\begin{eqnarray}  \label{DLP} \Delta x > l_P~. 
\end{eqnarray}

Notice that the considerations leading to (\ref{SQL}), (\ref{CGR})
and (\ref{DLP}) were in no way specific to an interferometer, and
hence are device independent. In summary, no device subject
to quantum mechanics, gravity and causality can exclude the quantization
of position on distances less than the Planck length.

\section{Minimal Length and Unimodular Gravity}

General relativity is a scaleless theory:
\begin{eqnarray} \label{GRact}
S_{GR}=\frac{1}{16 \pi G} \int d^4 x \sqrt{- g} R(g)
\end{eqnarray}
varying this action with respect to the metric $g^{\mu\nu}$ leads to the well-known Einstein equations. The action (\ref{GRact}) is invariant under general coordinate transformations and this may seem at odd with the notation of a minimal or fundamental length in nature. This may suggest that a quantum mechanical description of general relativity will fix the measure of Einstein-Hilbert action $\sqrt{- g} $ to some constant linked to the fundamental length. In that case one is led to unimodular gravity:
\begin{eqnarray} \label{GRactuni}
S_{GR}=\frac{1}{16 \pi G} \int d^4 x  R(g)
\end{eqnarray}
with the constraint $\sqrt{- g}=$ constant which implies that only variation of the metric which respect this contraint may be considered. This is basically the argument made by  Anderson and Finkelstein \cite{Anderson:1971pn} in favor of a unimodular theory of gravity.

There may be different ways to implement a minimal length in a theory, but we shall concentrate on one approach based on a noncommutative spacetime which indeed leads to a unimodular theory of gravity.
 Positing a noncommutative relation between e.g. $x$ and $y$ implies  $\Delta x \Delta y \geq |
\theta^{xy}|\sim l^2$, with $[\hat x, \hat y]=i \theta^{xy}$ and where $l$ is the minimal length introduced in the theory. This also implies that a spacetime volume is quantized $\Delta V \geq l^4$.

 One of the motivations to consider a noncommutative spacetime is that the noncommutative relations for the coordinates imply the existence of a minimal which can be thought of being proportional to the square root of the vacuum expectation value  of $\theta^{\mu\nu}$ i.e. $l_{\mbox{min}} \sim \sqrt{\theta}$.  If this length is fundamental it should not depend on the observer.  Assuming the invariance of this fundamental length, one can show that there is a class of spacetime symmetries called noncommutative Lorentz transformations\cite{Calmet:2004ii} which preserve this length.  It has  recently  been shown\cite{Calmet:2005qm}, that there are also  general coordinate transformations $\xi^\mu(\hat x)$ that leave the canonical noncommutative algebra invariant and thus conserve the minimal length:
\begin{eqnarray} \label{a}
[ \hat x^\mu, \hat x^\nu ]=i \theta^{\mu \nu},
\end{eqnarray}
where $\theta^{\mu\nu}$ is constant and antisymmetric. They are of the form:
$\xi^{\mu}(\hat x)=\theta^{\mu\nu} \partial_{\nu} f(\hat x)$, where $f(\hat x)$ is an arbitrary field.  The Jacobian of these restricted coordinate transformations is equal to one. This implies that the four-volume element is invariant: $d^{4}x^{\prime}=d^{4}x$. These noncommutative transformations correspond to volume preserving diffeomorphisms which preserve the noncommutative algebra. A canonical noncommutative spacetime thus restricts general coordinate transformations to volume preserving coordinate transformations. These transformations are the only coordinate transformations that leave the canonical noncommutative algebra invariant.  They form a subgroup of the unimodular transformations of a classical spacetime. 

The version of General Relativity based on volume-preserving diffeomorphism is known as the unimodular theory of gravitation\cite{UNI}.
Unimodular gravity here appears as  a direct consequence of spacetime noncommutativity defined by a constant antisymmetric $\theta^{\mu\nu}$. One way to formulate gravity on a noncommutative spacetime has been presented in refs.\cite{Calmet:2005qm}. Our approach might not be unique, but if the noncommutative model is reasonable, it must have a limit in which one recovers the commutative unimodular gravity theory in the limit in which $\theta^{\mu\nu}$ goes to zero. For small $\theta^{\mu\nu}$ we thus expect
\begin{eqnarray} \label{NCaction}
S_{NC}= \frac{-1}{16 \pi G} \int d^4x  R(g^{\mu\nu}) + {\cal O}(\theta),
\end{eqnarray}
where $R(g^{\mu\nu})$ is the usual Ricci scalar 
Once matter is included, one finds the following equations of motion:
\begin{eqnarray} \label{eqofmo}
R^{\mu\nu}-\frac{1}{4} g^{\mu\nu}R=-8 \pi G (T^{\mu\nu}-\frac{1}{4} g^{\mu\nu} T^\lambda_{\ \lambda})+ {\cal O}(\theta).
\end{eqnarray}
These equations do not involve a cosmological constant and the contribution  of vacuum  fluctuations automatically cancel on the right-hand side of eq.(\ref{eqofmo}).  As done in  e.g. ref.\cite{unruh} we can use the Bianchi for $R$ and the equations of motion for $T= -8 \pi G T_{\lambda}^{\ \lambda}$ and find:
\begin{eqnarray} 
D_\mu (R+T)=0
\end{eqnarray}
which can be integrated easily and give $R+T=-\Lambda$, where $\Lambda$ is an integration constant. It can then be shown that the differential equations (\ref{eqofmo})  imply
\begin{eqnarray} \label{einstein}
R^{\mu\nu}-\frac{1}{2} g^{\mu\nu}R-\Lambda g^{\mu\nu}=-8 \pi G T^{\mu\nu}-{\cal O}(\theta),
\end{eqnarray}
i.e. Einstein's equations\cite{Einstein:1916vd} of General Relativity with a cosmological constant $\Lambda$ that appears as an integration constant and is thus uncorrelated to any of the parameters of the action (\ref{NCaction}).  As we have shown, one needs to impose energy conservation and the Bianchi identities to derive eq.(\ref{einstein}) from eq.(\ref{eqofmo}). Because any solution of Einstein's equations with a cosmological constant can, at least over any topologically $R^4$ open subset of spacetime, be written in a coordinate system with  $g=-1$, the physical content of unimodular gravity is identical at the classical level to that of Einstein's gravity with some cosmological constant\cite{unruh}.

\section{Cosmological implications of spacetime quantization}

We now come to the link between a fundamental length and cosmology and rephrase the arguments developed in refs.\cite{Ng:1999tx,Ng:2003jk,Ahmed:2002mj,LamPred,sorkin2} within the framework of a fundamental length. It has been shown that the quantization of an unimodular gravity action proposed by Henneaux and Teitelboim\cite{henneaux}, which is an extension of the action defined in eq. (\ref{NCaction}),  leads to an uncertainty relation between the fluctuations of the  volume $V$ and those of the cosmological constant $\Lambda$: $\delta V \delta \Lambda\sim 1$ using natural units, i.e. $\hbar=l_p=c=m_p=1$. Now if spacetime is quantized, as it is the case for noncommuting coordinates, we expect the number of cells of spacetime to fluctuate according to a Poisson distribution, $\delta N \sim \sqrt{N}$, where $N$ is the number of cells. This is however obviously an assumption which could only be justified by a complete understanding of noncommutative quantum gravity. It is then natural to assume that the volume fluctuates with the number of spacetime cells $\delta V=\delta N$. One finds $\delta V \sim \sqrt{V}$ and thus  $\Lambda \sim V^{-\frac{1}{2}}$, i.e., we obtain an effective cosmological constant which varies with the four-volume as obtained in a different context in refs.\cite{Chen:1990jw,Ng:1999tx,Ng:2003jk,Ahmed:2002mj}. In deriving this result, we have assumed as in refs.\cite{Ng:1999tx,Ng:2003jk} that the fluctuation are around zero as explained below.  A minimal length thus leads leads to a vacuum energy density $\rho$
\begin{eqnarray} \label{crit}
\rho \sim \frac{1}{\sqrt{V}}.
\end{eqnarray}
Here we assume that the scale for the quantization of spacetime is the Planck scale. A crucial assumption made in refs.\cite{Ng:1999tx,Ng:2003jk,Ahmed:2002mj} as well is that the value of cosmological constant fluctuates around zero. This was made plausible by Baum\cite{Baum:1984mc} and Hawking\cite{Hawking:1984hk} using an Euclidean formulation of quantum gravity. 

Now the question is really to decide what we mean by the four-volume $V$. If this is the four-volume related to the Hubble radius $R_H$ as in  refs.\cite{Ng:1999tx,Ng:2003jk,Ahmed:2002mj} then this model predicts $\rho \sim (10^{-3} eV)^4$ which is the right order for today's energy density, it is however not obvious what is the equation of state for this effective cosmological constant. The choice $V=R_H^{4}$ might be ruled out because of the equation of state of such a dark energy model as shown in ref.\cite{Hsu:2004ri} in the context of holographic dark energy which leads to similar phenomenology. However if we assume that the four-volume is related to the future event horizon as suggested by M. Li\cite{Li:2004rb}, again in the context of holographic dark energy, then we get an equation of state which is compatible with the data $w=-0.903+1.04z$ which is precisely the equation of state for the holographic dark energy obtained in ref.\cite{Li:2004rb}. Details will appear in a forthcoming publication.

\section{Conclusions}
We have argued that an unification of quantum mechanics and general relativity implies that there is a fundamental length in Nature in the sense that no operational procedure would be able to measure distances shorter than the Planck length. Further we give an explicit realization of  an old proposal by Anderson and Finkelstein who had argued that a fundamental length in nature would imply unimodular gravity. Finally, using hand waving arguments we show that a minimal length might be related to the cosmological constant, which if this scenario is realized, is time dependent and thus only effectively a constant. Much more work remains to be done to establish this connection. It would be interesting to related the time dependence of the cosmological constant to that of other parameters of the standard model such as the fine-structure constant. Indeed as argued in refs.\cite{Calmet:2001nu} if one of the parameters of the standard model, such as a gauge coupling, a mass term or any other cosmological parameter,  is time dependent, it is quite natural to expect that the remaining parameters of the theory will be time dependent as well.

\section*{Acknowledgments}
I would like to thank Professor Xiao-Gang He and the Physics Department of the National Taiwan University for their hospitality during my stay at NTU. I am grateful to  Professors  Xiao-Gang He and Pauchy W-Y. Hwang for their invitation to present this work at the  CosPA 2006 meeting. This work was supported in part by the IISN and the  Belgian science policy office (IAP V/27).

\end{document}